\documentclass[12pt,psfig,epsfig]{article}
\usepackage{graphicx}
\usepackage{dcolumn}
\usepackage{bm}
\usepackage{epsfig}
\newcommand{\beq}{\begin{equation}}
\newcommand{\eeq}{\end{equation}}
\newcommand{\bea}{\begin{eqnarray}}
\newcommand{\eea}{\end{eqnarray}}

\def\ApJ{{\it ApJ} }
\def\ApJL{{\it ApJ} }
\def\ApJS{{\it ApJS} }

\def\AA{{\it A\&A} }

\def\MNRAS{{\it MNRAS} }
\def\Nature{{\it Nature} }

\def\PLB{{\it Phys. Lett.}{\bf B} }

\def\PRD{{\it Phys. Rev.} {\bf D} }
\def\PRL{{\it Phys. Rev. Lett} }
\def\RMP{{\it Rev. Mod. Phys.} }
\def\RPP{{\it Rep. Pro.Phys.} }

\begin{document}
\begin{center}
{The Galactic Wind Haze and its $\gamma$-spectrum
}\\
\medskip
{Nayantara Gupta$^{1}$, Biman B. Nath$^{1}$, Peter L.
Biermann$^{2,3,4,5,6}$,
Eun -Suk Seo$^{7}$,
Todor Stanev$^{8}$, Julia Becker Tjus$^{9}$}\vspace{0.4cm}\\
\parbox{\textwidth}{
$^{1}$ Raman Research Institute, Sadashiva Nagar, Bangalore 560080, India\\
$^{2}$ MPI for Radioastronomy, Bonn, Germany\\
$^{3}$ Dept. of Phys., Karlsruher Institut f\"ur Technologie, Karlsruhe,
Germany\\
$^{4}$ Dept. of Phys. \& Astr., Univ. of Alabama, Tuscaloosa, AL, USA\\
$^{5}$ Dept. of Phys., Univ. of Alabama at Huntsville, AL, USA\\
$^{6}$ Dept. of Phys. \& Astron., Univ. of Bonn, Germany \\
$^{7}$ Dept. of Physics, Univ. of Maryland, College Park, MD, USA\\
$^{8}$ Bartol Research Inst., Univ. of Delaware, Newark, DE, USA\\
$^{9}$ Institut f\"ur theoretische Physik, Plasma-Astroteilchenphysik, \\
$^{ }$    Ruhr-Universit\"at Bochum, D-44780, Bochum, Germany
}

%


\end{center}

\begin{abstract}
We study the possibility that the gamma ray emission in the Fermi bubbles observed is produced by cosmic ray electrons with a spectrum similar to Galactic cosmic rays. We argue that the cosmic ray electrons steepen near 1 TeV from $E^{-3}$ to about $E^{-4.2}$, and are partially secondaries derived from the knee-feature of normal cosmic rays. We speculate that the observed feature at $\sim 130$ GeV could essentially be due to inverse Compton emission off a pair-production peak on top of a turn-off in the $\gamma$ ray spectrum at $\sim 130$ GeV. It suggests that the knee of normal cosmic rays is the same everywhere in the Galaxy. A consequence could be that all supernovae contributing give the same cosmic ray spectrum, with the knee feature given by common stellar properties; in fact, this is consistent with the supernova theory proposed by Bisnovatyi-Kogan (1970), a magneto-rotational mechanism, if massive stars converge to common properties in terms of rotation and magnetic fields just before they explode.
\end{abstract}

\section{Introduction}
Fermi-LAT has revealed many interesting phenomena in the gamma ray sky.
The Fermi bubbles have been observed extending up to 10 kpc upwards and below from the Galactic Center (GC) \cite{su10} in the $1-100$ GeV energy range.
More recently Fermi group has reported the detection of a GeV line-like emission feature from the Galactic center region \cite{ackermann12}, which appears less significant in \cite{ackermann13}. The diffuse gamma ray data taken by Fermi-LAT appear to show a sharp feature, a line or lines from $1.5^{o}$ west of the Galactic Center (GC) \cite{su12}. 

The observational features of the line like structure are not yet clear. More sensitive detectors are needed to confirm and resolve this structure \cite{li12,mirabal13,hooper12} and then it would be possible to know whether these are lines and dark matter signatures or some astrophysical broadband effect. Su \& Finkbeiner \cite{su12} have interpreted the observational data as a single spectral line at $127 \pm 2$ GeV or a pair of lines at $110.8 \pm 4.4$ GeV and $128.8 \pm 2.7$ GeV giving a slightly better fit. The analysis of the Fermi-LAT data in \cite{cohen12} prefers a single line either at 130 GeV or 145 GeV. The line emission has been explained earlier with dark matter annihilations 
\cite{bringmann1,weniger12,tempel12,
kyae12,lee12,rajaraman12,buckley12,su12,hea12,
yang12}.
\par
Cosmic rays may also explain the gamma ray feature at 130 GeV. It has been suggested that a break in the gamma ray spectrum of the Fermi bubbles may appear as an excess at 130 GeV \cite{prof2012}.
\par
The interstellar radiation (IR) field has been calculated by \cite{moskalenko06} as a function of the Galactocentric radius, taking into account the emission by stars of different types and the scattering, absorption and re-emission of starlight by dust grains.
The IR energy spectrum has distinct peaks near $1\mu$m, $100\mu$m and $1000\mu$m. The peak at $1\mu$m or near $1 eV$ has the highest energy density.

 In the present work we have calculated the gamma ray spectrum produced by the Inverse Compton (IC) scattering of the multi-wavelength background photons by the cosmic ray electrons.
\par
 The break in the electron spectrum at 1 TeV where the electron spectrum steepens from $E^{-3}$ to about $E^{-4.2}$ produces a change in slope in the gamma ray spectrum near 130 GeV. Moreover, the threshold energy for pair production in $e\gamma$ interactions with 1 eV IR photons is close to 1 TeV. The electrons near the break at 1 TeV are expected to produce pairs and they subsequently emit gamma rays by IC scattering and possibly give rise to a feature at 130 GeV. We then show how this fits into a broader interpretation of recent cosmic ray data, and conclude that the underlying cosmic ray electron spectrum near TeV energies very likely is the same in the Fermi bubbles and near us. This strongly suggests that these electrons are partially secondary from spallation of cosmic ray nuclei near the knee energies, of about a few PeV; the cosmic ray nuclei spectrum then also has to be essentially the same at high energy in the Fermi bubbles and near us, and quite specifically show the knee feature at the same energy.
\section{Star Formation at Galactic Center and Fermi Bubbles}
Multi-wavelength observations have revealed surprising phenomena near the core of our Galaxy. The Fermi-LAT gamma ray detector has observed two large bubbles symmetrically located below and above the center of our Galaxy. The bubbles extend up to 10 kpc above and below the Galactic plane with a width of $40^o$.
\par
More interestingly, the WMAP \cite{jaro11} and PLANCK \cite{ade12} haze is located within the northern bubble sharing the same edges and the ROSAT soft X-ray maps \cite{snow97} are in the circumference of the bubbles. These observational correlations suggest a common origin of these emissions. The gamma ray energy flux in the energy range of $1-100$ GeV from Fermi bubbles is $E_{\gamma}^2 dN(E_{\gamma})/dE_{\gamma}=3\times10^{-7}$ GeV cm$^{-2}$ s$^{-1}$ sr$^{-1}$ and the solid angle subtended is $0.808$ sr \cite{su10}.

\par
This gamma ray emission has been explained as a result of cosmic ray protons interacting with protons in the medium \cite{crocker11} which leads to the production of charged and neutral pions.

A more detailed discussion on mass and energy flows through the Galactic Center and into the Fermi bubbles is given in \cite{crocker12}.

The bubbles are also explained as evidence of possible AGN jet activity in our Galaxy \cite{guo11}, collimation of a wide angle outflow from $Sgr A^*$ \cite{zubovas12} by the Central Molecular Zone.

\par
The magnetic field structure in the northern Fermi bubble has been explored with polarized microwave emission \cite{jones12}. Their study reveals that the magnetic field lines in the northern bubble's eastern wall and Galactic Center Spur are almost perpendicular to their extensions above the Galactic plane.
\par
We consider the case of a bubble being produced by the action of star formation in the Galactic Center region. Recent observations deduce a star formation rate of $\sim 0.04-0.08$ M$_\odot$ yr$^{-1}$, lasting about $\sim 1-10$ Myr \cite{yousef09,immer12}. The combined rate of momentum deposition from supernovae and stellar winds is then estimated to be $F \sim 5 \times 10^{33} (SFR/1 \, {\rm M}_{\odot}/{\rm yr}) \sim 2.5 \times 10^{32}$ dyne for a $SFR\sim 0.05$ M$_{\odot}$ yr$^{-1}$ \cite{leitherer99}. There is also likely to be a component of pressure from cosmic rays driving this wind \cite{everett08,uhlig12}. If we consider a bubble blown by the combined momentum injection traversing a distance of $L \sim10$ kpc in 15 Myr then the total energy deposited is $\sim F \times L \sim 7.7 \times 10^{54}$ erg. A fraction $\sim 15 \%$ of this energy going to cosmic rays would produce a cosmic ray energy budget of $\sim 10^{54}$ erg, consistent with the estimate of \cite{su10}. The implied energy density in a bubble of radius $\sim 5$ kpc is $\sim 1.4 \times 10^{-11}$ erg cm$^{-3}$, and the equipartition magnetic field is estimated to be $\sim 4 \, \mu$G, consistent with the recent measurement of magnetic field in the bubble by \cite{carretti13}.

The wind is unstable \cite{seemann97} and is traversed by many weak shocks, quite analogous to the winds of massive stars \cite{owocki84,owocki85,owocki88}; this analogy suggests the following interpretation: each shock raises the speed of sound a bit, and so each shock catches up with a previous shock. So ultimately, the speed of sound is so high, that a new shock no longer forms, and the combined energy of a large number of previous shocks is combined into one very large shock, that looks like a bubble.
\section{The cosmic ray spectrum in the Galactic Center region}
The spectrum of cosmic rays is determined by their injection spectrum, their transport processes, and their losses. In a series of papers \cite{biermann93,biermann93b,biermann93c,stanev93} it has been proposed, that the injection of Galactic cosmic rays can be divided into two main source classes, one due to explosions into the interstellar medium, and one due to explosions into the predecessor stellar wind. For the ISM-SN-CRs the prediction had been spectral index of $E^{-2.75\pm 0.04}$ after taking into account transport processes (with a Kolmogorov spectrum), $E^{-2.67+0.00-0.04}$ for wind-SN-CRs below the knee, and $E^{-3.07+0.00-0.14}$ above the knee. The quantitative predictions were summarized in \cite{biermann01}. The most recent fits to data give spectral index -3.08 beyond the knee 
\cite{biermann03,abbasi12a,abbasi12b,biermann12}.

The knee energy scales with the charge of the nucleus $\sim 600Z TeV$, with a large uncertainty. The wind-Supernovae also have a polar cap component with an injection spectrum of $E^{-2}$ and a sharp cutoff at the knee-energy due to a spatial limit in the acceleration region. The final cutoff for ISM-SN-CRs had been predicted to be around $100 \, Z$ TeV, and for wind-SN-CRs at about $100 \, Z$ PeV (see \cite{biermann94,biermann01}). This 
proposal has been shown to be supported by a number of observations, including
the CR-electron and CR-positron spectra \cite{biermann09,pamela}, the WMAP haze and 511 keV emission line \cite{biermann10a}, the spectra of the various elements constituting the ISM and the wind components \cite{biermann10b}, and the different spallation histories for ISM-SN-CRs and wind-SN-CRs \cite{nath12}, with a low interaction column for ISM-SN-CRs and a high column for the wind-SN-CRs. We note especially, that the new AMS \cite{aguilar13} and PAMELA data \cite{pamela}
 provide a precise test of the polar cap model; the data fit the prediction of \cite{biermann09} to better than two percent in spectral index. The spectrum of the cosmic rays beyond the knee was tested in \cite{biermann12}. \par
Here we address the possibility that in the observed gamma ray spectrum from the Fermi bubbles provide another test of this proposal.
For this we summarize below the properties of the relevant cosmic ray electrons.
After propagation the polar cap component of electrons from wind-SNe steepens to $E^{-2-1/3}$ due to diffusion losses and at higher energy $E^{-2-1}$ due to synchrotron and IC losses.
The polar cap component of $E^{-7/3}$ in the observer frame has a sharp cut-off at the knee energy, while the $4 \pi$ component goes from $E^{-8/3}$ to about $E^{-3.2}$ at exactly the same energy; the sum of these two components will therefore show a sharp drop at the knee energy, before continuing with the power-law $E^{-3.2}$. Pitch angle scattering and smoothing in the spallation production of the secondary leptons will modify this original spectral shape. In loss dominance over diffusion dominance this corresponds to a steep drop to a lower level power-law of $E^{-3.9}$. We approximate this combination of a sharp drop to a lower level $E^{-3.9}$ power-law with a simple power-law of $E^{-4.2}$, but note that depending on spectral resolution the effective power-law might be even steeper.
%
\begin{figure}
\vskip -1cm
\centering
\includegraphics[scale=0.7]{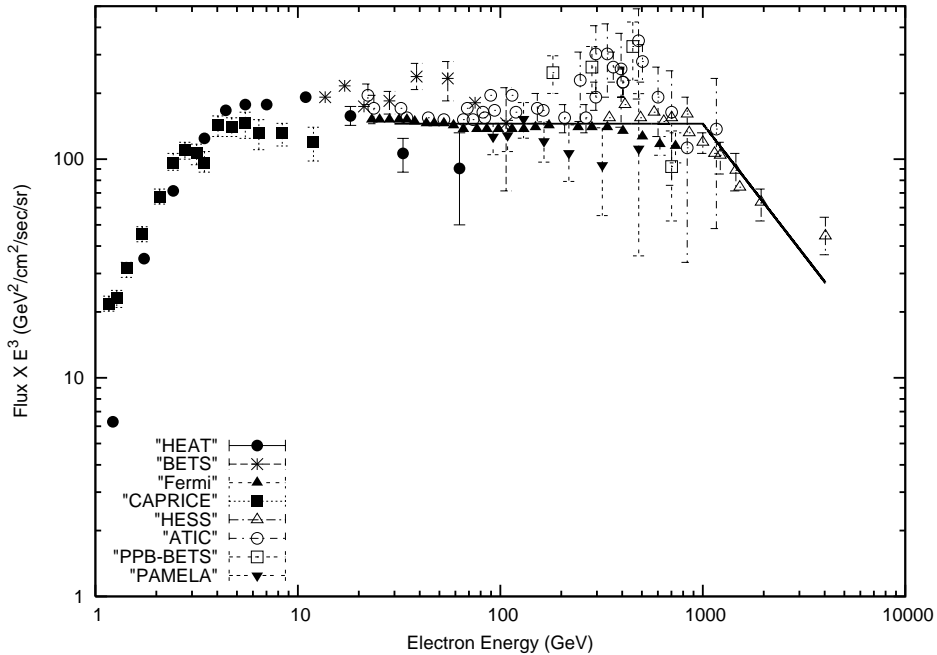}
\caption{ Observed electron spectrum near the Earth proportional to $E^{-3}$ and $E^{-4.2}$ below and above the break at 1 TeV respectively.}

\end{figure}

Interestingly, a break near 1 TeV in the electron spectrum has been observed near Earth \cite{biermann09} and it is expected to be present also near the Galactic Center region.
The knee near 4 PeV in the high energy cosmic ray spectrum can be related to the break in the electron spectrum at 1 TeV. Near the knee the cosmic rays are mostly He and heavier nuclei.
The Lorentz factor of the heavy nuclei at the knee $\gamma_{knee}=Z/A \gamma_p$, where $\gamma_p$ is the Lorentz factor of the protons. $Z/A$ is nearly $1/2$ for all heavy nuclei. We therefore find $\gamma_{knee}=2\times10^6$, which
is also the Lorentz factor of the 1 TeV cosmic ray electrons at the break.
The secondary electrons are perhaps produced in nuclear de-excitation following energetic particle interactions \cite{ramaty79} and as a result their Lorentz factors are approximately the same as that of the parent nuclei.
\par
The cosmic ray electron spectrum observed near Earth is shown in Figure 1: in the energy range of 100 GeV to 4 TeV. {We will show below that if the steepening from $E^{-3}$ to $E^{-4.2}$ is at 1 TeV then the IC scattering of the IR photons by the electrons at the break energy can explain the change in slope in the gamma ray spectrum near 130 GeV.
This requires the cosmic ray electron spectrum around TeV energies to be essentially the same in shape near the Galactic Center as near Earth.

\section{Gamma Ray Spectrum from Fermi Bubbles}
 The polar cap component of the electron spectrum
(in the loss dominant case) \cite{biermann09} as observed near the Earth is
\beq
\frac{dN_e}{d\gamma_e}  =  23.55 \times 10^{-6}  \gamma_e^{-3} cm^{-3}
\label{e_spec}
\eeq
We assume near the Fermi bubble region the electron flux is $\eta > 1$ times the flux observed near the Earth. 
This spectrum implies a cosmic ray energy density $\simeq 2\eta \times 10^{-12} erg/cm^{-3}$, consistent with that observed in the Fermi bubble region, as mentioned in section 2.
The $\gamma$ ray flux produced by IC scattering of IR photons is calculated.
\beq
E_{\gamma}^2  \frac{dN(E_{\gamma})}{dE_{\gamma}}  =  E_{\gamma} \frac{Vol}{4 \pi D_{GC}^2}  \frac{dE_{IC}}{dV dE_{\gamma} dt} \frac{1}{\triangle \Omega}
\label{gamma_energy_flux}
\eeq

We consider the region with $|b|\geq 30^{o}$ which is similar to the region of gamma ray observation in Fig.14. of \cite{su10}. The solid angle subtended by this region to the observer on the Earth is $\triangle \Omega=2\times\int_{0}^{180^{o}}d\phi\int_{0^o}^{60^o} \sin{\theta}d\theta=\pi sr$, where $\phi=l$ and $\theta=90^o-b$. The distance to GC from the Earth is $D_{GC}=7.5$ kpc 
\cite{nishiyama06,eisenhauer05} which gives $4\pi D_{GC}^2=10^{45.8}$ cm$^2$.
\par
The electron spectrum has a spectral index $p=3$ below the break at 1 TeV
and above it is $p=4.2$.
The expression for power emitted in IC scattering by electrons is
(eqn(7.28a) of \cite{rybicki79}
\beq
\frac{dE_{IC}}{dV dE_{\gamma} dt} =  \frac{3}{4} C  \eta  c 
\sigma_{T} \int\Big(\frac{E_{\gamma}}{\epsilon}\Big) v(\epsilon) d\epsilon\int_{\gamma_{e1}}^{\gamma_{e2}} {\gamma_e}^{-p-2}f\Big(\frac{E_{\gamma}}{4{\gamma_e}^2\epsilon}\Big) d\gamma_e
\label{IC_power}
\eeq
where $C=23.55 \times 10^{-6}$, $v(\epsilon)$ is the low energy photon density per unit energy $\epsilon$, and $c$ is the speed of light.
We have used the full IR background spectrum from \cite{moskalenko06}.
The functional form of $f\Big(\frac{E_{\gamma}}{4{\gamma_e}^2\epsilon}\Big)$ in the Thomson regime has been given in \cite{rybicki79}. In the Klein Nishina regime this function has an extra term as given in \cite{jones68,georgan04}.
We have taken $\eta \times Vol/(10^{66} \, {\rm cm}^3)\sim 6$, where $\eta$ is taken to be of order unity in order to take into account the possible variation in CR flux in the galactic center area, and also the uncertainty in our knowledge of the volume.
\par
For $\gamma_e \epsilon\geq 0.2 m_e c^2$ Klein Nishina effects become significant in IC scattering and we have included it in our calculation following the formalism discussed in \cite{georgan01,georgan04}. 
In our case the collisions of the electrons and the incoming low energy photons are isotropic in the observer's frame and this has been used of in our calculation of gamma ray flux. The flux of the gamma rays is calculated using eqn(\ref{gamma_energy_flux}), eqn(\ref{IC_power}) and compared with the observed gamma ray flux from Fermi bubbles from \cite{su10} in Figure 2 of this paper.

\section{Pair Production in $e\gamma$ Interactions}

It is important to note that very high energy electrons lose energy by producing $e^{-}e^{+}e^{-}$ in collisions with IR photons \cite{blumenthal70}. The electron positron pairs produced in $e\gamma$ collisions produce more pairs after colliding with the background IR photons. The energy of the electrons and IR photons in the centre of mass(CM) frame and observer's frame are related as
\beq
E_{CM}=[m_e^2c^4+2E_e\epsilon(1-\beta_ecos{\theta})]^{1/2}
\label{energy_transf}
\eeq
where $\beta_e\sim1$, $\theta$ is the angle between the colliding electron and IR photon in the observer's frame, for isotropic collisions the average value of $\theta$ is $\pi/2$. The threshold energy for pair production in $e\gamma$ interaction is $E_{th,CM}=3m_e c^2$ in the CM frame. It follows from eqn(\ref{energy_transf}) that the minimum energy of the incoming electron in the observer's frame is
\beq
E_{e}=\frac{8m_e^2c^4}{2\epsilon(1-\beta_e cos{\theta})}
\label{electron_en}
\eeq
With $\epsilon=1$eV we find the minimum energy of the parent electron has to be $E_e=1.044$TeV in the observer's frame. This is close to the break energy in the electron spectrum at $1$ TeV. 
\par
If the observed gamma ray emission is isotropic then the $e^{+}e^{-}e^{+}$ produced in $e\gamma$ collisions are expected to be emitted isotropically.
The CM frame has total momentum zero. It is at rest with respect to the observer's frame as a result we also expect isotropy in this frame.  
Momentum conservation in CM frame allows us to obtain the ratios of the energies of the electrons and positrons produced in $e\gamma$ collisions.
For isotropic emission of the leptons the average angle between them has to be 
$2\pi/3$. With this angle we find the ratio of the momenta of the leptons is 
$1:1:1$. They share the parent electron's energy equally, so each of them has 
 energy $1044/3=348 GeV$ in observer's frame.  
 
The IR photons are IC scattered off the daughter leptons. The energy of the gamma rays produced in IC of the pairs is calculated using eqn.(7.2) of \cite{rybicki79}, after transforming it to observer's frame from electron rest frame.
\beq
E_{\gamma}=
\frac{\gamma_e^2\epsilon}{1+\frac{\gamma_e\epsilon}{m_e c^2}(1-cos\phi)}
\label{IC_en}
\eeq
 $\phi$ is the angle of emission of the scattered photon in the electron's rest frame. In the observer's frame the collisions and scatterings are isotropic with average angle of emission $\phi_{obs}=\pi/2$. 
 The Lorentz transformation of angles between the observer's frame and electron's rest frame has been used following eqn(4.8a) of \cite{rybicki79}.
\beq
tan\phi_{obs}=\frac{sin\phi}{\gamma_e(cos\phi+v_e/c)}
\eeq
where, $\gamma_e= 7\times 10^5$ is the Lorentz factor of the electron and $v_e\sim c$ is its velocity.  We find in electron's rest frame the average value of $\phi$ is $\pi$.
The $\gamma$ ray emission is highly Lorentz boosted in one direction in the 
lepton's rest frame.  The energy of the gamma rays produced in IC emission of $1 eV$ IR photons off the $348GeV$ electrons and positrons is 126 GeV in the observer's frame. These gamma rays may contribute to the line feature observed near 130 GeV. 
The width is given by 
three effects, a) the spread between threshold and cutoff, b)  the 
angular spread allowed by staying above threshold in CM frame, and c) 
the small fraction of slightly different secondary lepton energies.

\begin{figure}
\centering
\includegraphics[bb=0cm 0cm 21.6cm 28.0cm,viewport=4.6cm 15cm 18.0cm 26.0cm,clip,scale=0.8]{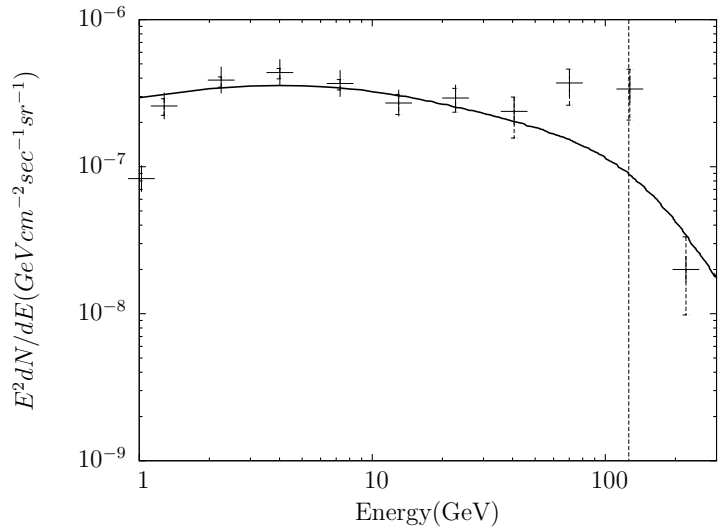}
\vskip -1cm
\caption{ Gamma ray data from the bubbles with error bars from Fig.14 of 
\cite{su10}.}
\end{figure}

\section{Discussions}
We have described a model to explain the gamma ray emission observed by Fermi-LAT from GC region and extending upto 10 kpc as IC emission of cosmic ray electrons. In our model, it is the IR photons in this region are IC scattered off the cosmic ray electrons. We have also included the microwave photons in our calculation but their contribution is less compared to the IR photons.

\par
The parent electrons of energy above $1.044 TeV$ satisfy the threshold condition for pair production with the IR photons at 1 eV where the IR background has a peak. The electron positron pairs produced in $e\gamma$ collisions have average energy $348 GeV$. The IC scattering of IR photons of energy 1 eV off these secondary pairs leads to the production of gamma rays of average energy $126 GeV$. 
These gamma rays on top of the bending in the gamma ray spectrum near $130 GeV$ due to the break in the electron spectrum at 1 TeV sharpen the line feature.

\par
The Galactic wind driven by star formation in the GC region and the emitted cosmic rays supplies the energy of $\sim 10^{54}$ erg to the bubbles of gamma rays. 
A crucial aspect of our model is that the break in the bubbles' spectrum is due to the break in the spectrum of cosmic ray electrons, which is similar to that
 found near the Earth.
\par
We discuss below a few aspects of this scenario.
The intensity of gamma rays from the bubbles does not show any significant variation with distance from GC. This may be understood as follows: a) The gamma ray intensity of Fermi bubbles depends on the intensity of the background IR radiation field and the density of the cosmic ray electrons. The IR radiation is essentially constant initially up into the halo as long as the lateral scale of the emitting region is larger than the distance above the disk.
This follows from a result in electrostatics that for a lateral distribution of sources the vertical field is approximately constant up to a length scale which corresponds to the lateral extension of the source distribution.
b) The density of the cosmic ray electrons is inversely proportional to the cross-section of the flow. If the initial flow is straight up, then the density changes only slowly and adiabatic losses may be weak. Moreover, re-acceleration by weak shocks (eqn. 2.45 in \cite{drury83}) may keep the spectral shape of the cosmic ray electrons intact but sharpen the kink in the spectrum.
 c) What we observe is a line of sight integral, so even a decrease of the IR radiation field weaker or similar to $r^{-1}$ would be compensated in the integral for a straight initial flow up/down. Taken together this may allow to understand the constancy of the emission.
\par
Fermi-LAT has also observed emission of gamma rays in $1-100$ GeV band in the star forming region of Cygnus X \cite{ackermann11}. The observed gamma ray spectrum is proportional to $E_{\gamma}^{-2}$. The polar cap component of cosmic rays at the source is proportional to $E^{-2}$ and it gives secondary pions in hadronic interactions. The neutral pions subsequently decay to gamma rays.
Thus the gamma ray spectrum observed from Cygnus X might have originated in hadronic interactions of the polar cap component of cosmic rays. IC emission spectrum of secondary electrons in this case would be proportional to $E_{\gamma}^{-3/2}$ and inconsistent with the observed spectrum $E_{\gamma}^{-2}$.

\par
The direct observations of the cosmic ray electron spectrum near the Earth give us the loss-dominant spectrum of the cosmic ray electron spectrum in our 
neighbourhood, and the gamma-ray observations give us that spectrum in the Galactic Center region. It is very interesting that the derived spectral shape is the same as that observed near the Earth.
\par

It is possible to relate the knee in the very high energy cosmic ray spectrum with the break in the cosmic ray electron spectrum. Cosmic ray interactions and subsequent nuclear de-excitation leads to the production of the secondary electrons, as the cosmic ray parent nuclei and secondary electrons have the same Lorentz factor.
This implies that the cosmic ray knee is at the same energy in terms of E/Z everywhere in the Galaxy as the break energy in the electron spectrum is same in different regions of the Galaxy, as predicted from a theory, that attributes this kink to the original stars. This derives from the fact, that in different parts of the Galaxy very different stars contribute to the kink feature of cosmic rays, the knee, and in each location many stars contribute. So to have a clear feature at all, and then to have that feature the same in different parts of the Galaxy puts very strong constraints on the exploding stars: they must all be asymptotically similar at the point of explosion.
\par
We need to ask, whether any other mechanism could produce such a kink in the cosmic ray spectrum, and also give a break energy which is the same everywhere in the Galaxy: First we may consider the possibility that OB super-bubbles produce the cosmic ray component giving the knee; 
in such a theory a curvature upwards and below the energy of the knee would be consistent with arguments on shock structure (e.g. \cite{drury11}). One might speculate that it could also produce a knee from a characteristic length scale of the super-bubble. However, it is hard to see how this critical length scale can be the same everywhere in the Galaxy. Second, we can generalize this argument to the transport in the Galaxy, and the escape from the Galaxy. However, as is well known, this possibility would give a much larger cosmic ray anisotropy in arrival directions than what has been observed, and so is also wrought with difficulties (e.g. \cite{biermann93}, and the references mentioned there).
\par
So, it appears that the original exploding stars are required to explain this kink, called the knee. This has actually been predicted \cite{biermann93}, and is based on the magneto-rotational mechanism for the explosions of very massive stars proposed by \cite{bisnovatyi70}, and worked out in much more detail by 
\cite{ardeljan05,moiseenko06,bisnovatyi07} and others. This mechanism connects the rotation of the core of the star with the magnetic fields, just as is required to explain a constant knee E/Z scale \cite{biermann93}.

\par

We can also make further tests of the proposal. If the $E_e^{-3}$ segment of the
electron spectrum is partially secondary, then the upturn should be connected to the parallel upturn in the positrons (e.g. \cite{biermann09,adriani11}); in fact this is consistent, as already noted above, with the new AMS 
\cite{aguilar13} and PAMELA data \cite{pamela}.

The cosmic ray electrons also emit synchrotron photons
which have been observed in the PLANCK spectrum \cite{ade12}.
One zone compatible with radio spectrum $\nu^{-1}$ due to the electron spectrum $E_e^{-3}$ in the loss limit and another with a spectrum compatible with $\nu^{-2/3}$ due to the electron spectrum $E_e^{-7/3}$ in the diffusion limit are expected to be observed by  PLANCK. A ring ought to be present at the transition region between the outer zone of high frequency radio spectrum proportional to $\nu^{-1}$ and inner zone where it is proportional to $\nu^{-2/3}$ .
Perhaps due to ignoring the ring region the  PLANCK data is at present consistent with a slightly steeper spectrum in the outer zone and a flatter spectrum in the inner zone. It is also expected that there should be many wind-SN-remnants with a spectrum of $E^{-7/3}$ at lower  energies, and $E^{-2}$ at higher energies.

\section{Conclusions}

In this work we propose the gamma ray emission from the Fermi bubbles is the IC emission of the FIR radiation field by cosmic electrons/positrons produced in spallation of cosmic ray nuclei at the knee. The break in the electron spectrum near 1 TeV could be the origin of the change in slope in the gamma ray spectrum near 130 GeV. Moreover, the electrons near the break produce pairs in $e\gamma$ 
interactions which may contribute to the gamma ray spectrum, make the bend look sharper, and so could make it appear as a line feature from noise. 
We realize that this implies that the cosmic ray electron spectrum has the same shape at high energy in the GC region as near to Earth in the Galaxy. This implies then in our concept that the cosmic ray particles themselves have the same knee energy through out the Galaxy. This in turn supports an origin of this feature in the original supernova explosions, which in turn supports the magneto-rotational mechanism proposed by \cite{bisnovatyi70}.

If all this is true, we have identified a common quantitative feature of all very massive stars in the final stage just before they blow up, connecting rotation and magnetic fields.

\section{Acknowledgements}
PLB acknowledges discussions about this and related topics with W. Cui, R. Diehl, J. Finley, A. Haungs, A. Meli, A. Obermeier, M. Roth, D. Ryu, V. de Souza, and many others.
Support for E.-S.S. comes from NASA grant NNX09AC14G and for T.S. comes from DOE grant UD-FG02-91ER40626.

\end{document}